\documentclass[pra,aps,twocolumn,superscriptaddress,a4paper,showpacs]{revtex4}

\usepackage{graphicx,amssymb,amsmath}

\newcommand{\bra}[2][]{#1\langle #2 #1\rvert}
\newcommand{\ket}[2][]{#1\lvert #2 #1\rangle}

\newcommand{\Tr}[0]{\ensuremath{\mathrm{Tr}}}
\newcommand{\down}{\ensuremath{\downarrow}}
\newcommand{\up}{\ensuremath{\uparrow}}

\begin{document}
\title{Entanglement of indistinguishable particles in condensed matter physics}

\author{Mark~R.~Dowling}
\email{dowling@physics.uq.edu.au} \affiliation{School of Physical
Sciences, The University of
  Queensland, Queensland 4072, Australia}

\author{Andrew~C.~Doherty}
\email{doherty@physics.uq.edu.au}
\affiliation{School of Physical Sciences, The University of
  Queensland, Queensland 4072, Australia}

\author{Howard~M.~Wiseman}
\email{h.wiseman@griffith.edu.au}
\affiliation{Centre for Quantum
Computer Technology, Centre for Quantum Dynamics, School of Science,
Griffith University, Brisbane 4111, Australia}

\date{\today}

\begin{abstract}
  The concept of entanglement in systems where the particles are
  indistinguishable has been the subject of much recent interest and
  controversy.  In this paper we study the notion of entanglement of
  particles introduced by Wiseman and Vaccaro~[\prl {\bf 91}, 097902
  (2003)] in several specific physical systems, including some that
  occur in condensed matter physics.  The entanglement of particles is
  relevant when the identical particles are itinerant and so not
  distinguished by their position as in spin models.  We show that
  entanglement of particles can behave differently to other approaches
  that have been used previously, such as entanglement of modes
  (occupation-number entanglement) and the entanglement in the
  two-spin reduced density matrix.  We argue that the entanglement of
  particles is what could actually be measured in most experimental
  scenarios and thus its physical significance is clear.  This
  suggests entanglement of particles may be useful in connecting
  theoretical and experimental studies of entanglement in condensed
  matter systems.
\end{abstract}


\maketitle

\section{Introduction}

Recently there has been much interest in understanding and quantifying
the entanglement present in quantum many-body systems.  The aim of
this program of research is to shed new light on systems, particularly
strongly correlated systems, that are difficult to treat with
conventional approaches.  Most studies have focused on quantum spin
systems, especially near quantum phase transitions, see for example
\cite{osborne2002, osterloh2002, vidal2003, latorre2003a,
  verstraete2004, popp2005} and references therein.  The concept of
entanglement is well-defined in these systems as the spins can be
considered distinguishable.

The subject of entanglement becomes more subtle when the system to be
studied consists of many \textit{indistinguishable} particles.
Examples of such systems span many fields of physics: quantum optics
experiments, ultracold atomic gases, itinerant electrons and
superconductors. Even the question of which states are entangled is
the subject much recent debate~\cite{tan1991, hardy1994,
  greenberger1995, vanenk2005, paskauskas2001, schliemann2001, li2001,
  gittings2002, vedral2003, shi2003, wiseman2003, kaplan2005}. The
difficulty arises from the lack of individual identity of the
particles that are supposed to be entangled, which is manifest as the
necessary symmetrization or antisymmetrization of the quantum
wavefunction.

Wiseman and Vaccaro~\cite{wiseman2003} have recently proposed a
measure of entanglement for systems of indistinguishable particles
that is operational in the sense that it quantifies the amount of
``accessible'' entanglement in the system where a local particle
number superselection rule restricts the possible operations that may
be performed.  This is in contrast to other measures such as the mode
entanglement, where the physical meaning of the entanglement measure
is not so clear as the measurements required to demonstrate
entanglement are not obviously possible.  The rapidly-developing field
of mesoscopic electronics may provide a useful testing ground for
comparing different notions of entanglement in condensed matter
systems as experiments to demonstrate entanglement may be feasible in
the near future~\cite{samuelsson2003, beenakker2003, samuelsson2004,
  beenakker2005, samuelsson2005, samuelsson2006}.

The accessible entanglement, which is referred to as the
``entanglement of particles'' in~\cite{wiseman2003}, is defined as the
amount of entanglement that could be extracted from the system and
placed in conventional quantum registers, from which it could be used
to perform quantum information processing tasks, such as
teleportation. In many physical systems it may be difficult to extract
all, or even some, of this entanglement, but the quantity itself may
still give insight into the physical properties of the system.  An
analogy with thermodynamics is helpful.  In thermodynamics it is often
fruitful to consider quantities such as the amount of free energy in a
system, even in the absence of an explicit scheme to extract that free
energy.  If the total entanglement is taken to be analogous to the
total internal energy, then the accessible entanglement is somewhat
analogous to the free energy.

In~\cite{wiseman2003} the entanglement of particles was defined and
evaluated for a number of states of indistinguishable particles.
However there is a lack of studies of entanglement of particles in
explicit physical models.  In this work we aim to fill this gap by
investigating the entanglement of particles in a number of simple
physical systems, and compare and contrast to other approaches to
studying entanglement.

We begin in Sec.~\ref{sec:review} by reviewing the definition of
entanglement of particles and explaining its motivation in terms of
superselection rules and measurements. In Sec.~\ref{sec:smallsystems}
we study ground and thermal states of systems of bosons and fermions
with a small number of modes.  We show that for any number of modes
non-interacting bosons have zero entanglement of particles, as one
might intuitively expect. However non-interacting fermions can have
non-zero entanglement of particles.  We then study the effect of
interactions on the entanglement of particles using the Bose-Hubbard
and Fermi-Hubbard models as examples.  We contrast the behaviour of
the entanglement of particles with the entanglement of modes and show
that the two measures can display opposite behaviour as one varies the
interaction parameter.

In Sec.~\ref{sec:multimode} we turn to multimode systems.  In
\ref{sec:EPcorrel} we show how to calculate the entanglement of
particles from correlation functions.  A particularly striking example
of entanglement of indistinguishable particles is the non-interacting
electron gas as studied in~\cite{vedral2003,oh2004}.  In those works
the ``two-spin reduced density matrix'' --- a concept common in
many-body physics --- is used to study entanglement.  It is shown that
there is a finite length over which the non-interacting electrons are
entangled.  In \ref{sec:freeelectrons} we show a similar effect in a
lattice model of non-interacting electrons, where the entanglement
persists over many lattice sites.  In \ref{sec:continuum} we make the
connection to the continuum explicit and argue that writing down a
two-spin reduced density matrix on the lattice leads to difficulties
in interpreting the entanglement.  The subtle difficulty stems from
the indeterminate number of particles at any particular location.  Our
results show that the phenomenon of entanglement of non-interacting
electrons may feasibly be observed in an experiment.

\section{Superselection rules, measurements and accessible
entanglement}

\label{sec:review}

In this section we review the concept of entanglement of particles, as
defined by Wiseman and Vacarro~\cite{wiseman2003} and explain why we
consider it an appropriate measure of entanglement in condensed matter
systems.

In many-body physics it is common to represent the state of a system
in the \textit{occupation-number representation}.  If $\{ \psi_j \}$
is a complete set of single-particle wavefunctions (for example modes
localized in position or momentum) then a many particle state is
written
\begin{equation}
\ket{\Psi} = \sum_{\vec{n}} c_{\vec{n}} \ket{\vec{n}},
\end{equation}
where $\vec{n} = (n_1,n_2, \ldots)$ is a set of occupation numbers for
the single particle modes (for fermions the occupation numbers are
restricted to be 0 or 1 due to the Pauli exclusion principle), and the
$c_{\vec{n}}$ are coefficients in the superposition.  Formally, the
space of occupation-number states is equivalent to a tensor product
space where each mode is a factor (subsystem), and the occupation
number of each mode represents a distinct state in that subsystem. It
is thus tempting to define the ``entanglement'' in a many-body state
as being with respect to this mode decomposition.

Following~\cite{wiseman2003}, for bipartite systems we may quantify
the \textit{entanglement of modes}, $E_M$, as
\begin{equation}
E_M(\rho_{AB}) = M(\rho_{AB})
\end{equation}
where $M$ is some bipartite measure of entanglement (e.g.\ 
entanglement of formation, entanglement of distillation, negativity),
$A$ and $B$ each control some subset of the total modes, and
$\rho_{AB}$ is the total state shared by $A$ and $B$.  This approach
is advocated in, for example, \cite{zanardi2002, shi2003, hines2003a}.
The entanglement of modes depends on which modes $A$ and $B$ control,
as discussed in~\cite{zanardi2001,vanenk2003}, but not on the local
mode decomposition that they choose.

The difference between entanglement of particles, which we define
shortly, and entanglement of modes stems from the local
particle-number superselection rule which may apply to systems of
massive particles. Operationally, a superselection rule (SSR) is a
restriction on the allowed physical operations (closed or open
evolution, preparation, measurement, etc.) on a
system~\cite{bartlett2003}.

It is sometimes asserted that certain superselection rules apply
\textit{in principle} due to some underlying symmetry of the system,
e.g.\ a SSR for charge that appears in Lorentz-invariant quantum field
theories. However it is possible to lift superselection rules by
constructing an appropriate reference frame for the quantity in
question, the most famous example of this procedure being the thought
experiment of Aharonov and Susskind~\cite{aharonov1967}.

Whether or not superselection rules apply in principle is not
important for our purposes.  We simply note that often a
superselection rule applies in practice due to the lack of an
appropriate reference frame.  The example we will be concerned with is
the superselection rule for \textit{local} particle number.  Consider
a bipartite state of one particle superposed over two modes, where
each party controls one of the modes:
\begin{equation}
\ket{\psi_\theta}_{AB} = (\ket{1}_A \ket{0}_B + e^{i \phi} \ket{0}_A
\ket{1}_B)/\sqrt{2}.
\end{equation}
The phase $\phi$ in this superposition is only meaningful relative to
some shared reference frame.  Two examples of systems that could act
as a reference frame for this phase are a large coherent state of
light if the particle were a photon, or a Bose-Einstein Condensate if
it were a bosonic atom.  However without such a reference frame, as is
generally the case in condensed matter systems, the phase is not
accessible to experiment and the state $\ket{\psi_\theta}$ is
indistinguishable from the averaged state
\begin{eqnarray*}
\bar{\rho}_{AB} &=& \int_0^{2 \pi} \frac{d \theta}{2 \pi}
\ket{\psi_\theta}_{AB} \bra{\psi_\theta} \\
&=& (\ket{1}_A \bra{1} \otimes \ket{0}_B \bra{0} + \ket{0}_A \bra{0}
\otimes \ket{1}_B \bra{1})/2,
\end{eqnarray*}
which is an incoherent mixture of the particle being in one mode or
the other.

Notice that in the above example averaging over the unknowable phase
$\phi$ is equivalent to projecting onto fixed local particle number.
This is a general result --- if two parties, $A$ and $B$, share a
multimode state of indistinguishable particles, $\rho_{AB}$, and a
local particle number superselection rule applies, then this state is
indistinguishable from the averaged state
\begin{equation}
\bar{\rho}_{AB} = \sum_{n_A,n_B} \Pi_{n_A,n_B} \rho_{AB}
\Pi_{n_A,n_B},
\end{equation}
where $\Pi_{n_A,n_B}$ projects onto fixed particle number $n_A$ and
$n_B$ at $A$ and $B$, and the sum runs over all possible local
particle numbers~\cite{wiseman2003}.

For these reasons Wiseman and Vaccaro~\cite{wiseman2003} argue that
the entanglement of modes does not capture the true amount of
entanglement that the two parties, $A$ and $B$, share since in order
to take advantage of it they would need to be able to perform
arbitrary local operations on the modes.  In general such local
operations would violate the local particle number superselection rule
and are hence not possible in practice.

Wiseman and Vaccaro give an operational definition of bipartite
entanglement of indistinguishable particles by using the concept of a
\textit{standard quantum register} --- a set of distinguishable
qubits~\footnote{The qubits could be distinguished, for example, by
  their fixed positions in space.} --- which each party possesses.
They define the \textit{entanglement of particles} as the maximal
amount of entanglement that the two parties can produce between their
standard quantum registers by local operations on the modes that they
have access to. Because the standard quantum registers consist of
distinguishable qubits their entanglement may be measured by any
standard measure of bipartite entanglement.

In~\cite{wiseman2003} only pure states are considered, however a
definition of entanglement of particles $E_P$ that applies for mixed
states as well is
\begin{equation}
\label{eq:EPdef} E_P(\rho_{AB}) = \sum_{n_A,n_B} P_{n_A,n_B}
E_M(\rho^{(n_A,n_B)}_{AB}),
\end{equation}
where $\rho^{(n_A,n_B)}_{AB} = \Pi_{n_A, n_B} \rho_{AB} \Pi_{n_A,
  n_B}$ is the (unnormalized) state conditioned on obtaining the
results $n_A$ and $n_B$ for a measurement of local particle number at
$A$ and $B$, $P_{n_A, n_B} = \Tr(\rho^{(n_A,n_B)}_{AB})$ is the
probability of obtaining that result and $E_M(\rho^{(n_A,n_B)}_{AB})$
is the entanglement of modes in $\rho^{(n_A,n_B)}$~\footnote{We first
  normalize the state, $\rho^{(n_A,n_B)}_{AB}/P_{AB}$, before
  calculating the standard bipartite measure of entanglement, $M$.}.
In words, the entanglement of particles is the weighted sum of the
entanglement of modes when local particle number is measured. It is
sensitive, for example, to entanglement in spin between two particles
at distinct spatial locations, but not to ``occupation-number
entanglement'' such as exists mathematically in the state
$\ket{\psi_\theta}$ above but would be impossible to extract in an
experiment without a shared reference frame.

It would be most in the spirit of the operational definition to use
the \textit{distillable entanglement}\cite{bennett1996} as the
entanglement measure, $E_M$.  However this measure is often difficult
to calculate, so throughout this paper we use \textit{entanglement of
  formation} instead, as it is generally easier to calculate.  In
general the distillable entanglement is less than the entanglement of
formation, and it is possible for a quantum state to have non-zero
entanglement of formation but zero distillable entanglement.

The effective measurement of local particle number that appears in the
definition of entanglement of particles is formally due to a lack of
phase reference, as discussed above.  However in many experimental
scenarios the measurement does actually occur. For example in
measuring correlations in spin between electrons in a mesoscopic
conductor a measurement of spin (up or down) simultaneously implies
that an electron was also measured at the location of the detector.
In measuring a mode or modes, e.g.\ momentum mode/s in a mesoscopic
conductor, the Hamiltonian coupling the measuring apparatus to the
mode will typically commute with the total occupation number for the
mode/s. It is straightforward to show that under these circumstances
the set of generalized measurements (POVMs)~\cite{nielsen2000a} that
can be implemented have Kraus operators that commute with the total
occupation number. This implies that the generalized measurements obey
the superselection rule.  An example of a measurement that does not
commute with local particle number, and therefore does not obey the
superselection rule, is a projective measurement in the basis $\{
(\ket{0}+\ket{1})/\sqrt{2}, (\ket{0}-\ket{1})/\sqrt{2} \}$.

Practically it is often convenient to use entanglement witness to
prove that a certain state is entangled.  We note that entanglement of
particles could be detected by measuring an entanglement witness that
commutes with local particle number, as discussed for optical lattices
in~\cite{toth2004}.  In other words states which are entangled in
modes but not in particles are not detected by this type of
entanglement witness.

\section{Some simple systems}

\label{sec:smallsystems}

We now study some simple systems using the entanglement of particles
in order to build intuition for its behaviour before moving to
multimode systems in the next section.

In order to share a state with non-zero entanglement of particles $A$
and $B$ must each be in control of at least two modes, and there must
be at least two particles in the system.  Therefore the simplest
possible system in which there is entanglement of particles is two
particles in four modes.

In these minimal systems the entanglement of particles is due solely
to entanglement of the modes at $A$ and $B$ when there is one particle
at each location (i.e.\ only the $n=1$ term from Eq.~(\ref{eq:EPdef})
contributes). Because there are two modes at $A$ and $B$ we have an
effective two qubit system for which it is possible to calculate the
entanglement of formation in closed form as a function of the density
matrix for pure~\footnote{For pure states it is equal to the
  distillable entanglement, and given by the von Neumann entropy of
  the reduced state of either party.} or mixed
states~\cite{wootters1998}.

\subsection{Two bosons in four modes}

Perhaps the simplest model for interacting bosons on a lattice is the
Bose-Hubbard model.  For four lattice sites (modes) the Hamiltonian is
\begin{equation}
\label{eq:bhfourmode} \hat{H} = -t \sum_{j=0}^3 (\hat{b}^\dagger_{j}
\hat{b}_{j+1} + \hat{b}^\dagger_{j+1} \hat{b}_j) + U \sum_{j=0}^3
\hat{n}_{j} (\hat{n}_{j} - 1),
\end{equation}
where $\hat{b}_j$, $\hat{b}^\dagger_j$ are the usual boson
annihilation and creation operators that satisfy
$[\hat{b}_j,\hat{b}_{j'}^\dagger] = \delta_{j, j'}$, $\hat{n}_j =
\hat{b}^\dagger_j \hat{b}_j$ is the number operator for site $j$, and
we have imposed periodic boundary conditions, $j+1 = 0$ for $j = 3$.

\begin{figure}
\centering
\includegraphics[]{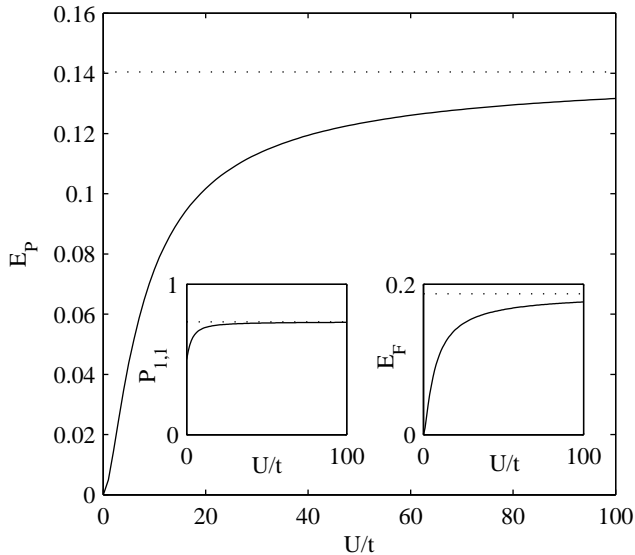}
\caption{Bipartite entanglement of particles in the ground state of
  two bosons in four modes with Bose-Hubbard Hamiltonian and periodic
  boundary conditions.  $A$ and $B$ control adjacent modes in the
  ring. $E_P$ is zero at $U=0$ (non-interacting bosons) and saturates
  at $(3/4) h(1/2+\sqrt{2}/3) \simeq 0.1405$ for large $U/t$
  corresponding to $P_{1,1} = 3/4$ and $E_F = h(1/2+\sqrt{2}/3)$.}
\label{fig:2bosons4modesground}
\end{figure}

For a fixed total number of bosons, $N=2$ say, we write out the
Hamiltonian matrix in the Fock basis and calculate the eigenvalues and
eigenstates.  Fig.~\ref{fig:2bosons4modesground} shows the
entanglement of particles in the (non-degenerate) ground state of
Eq.~(\ref{eq:bhfourmode}) as a function of $U/t$, where both $A$ and
$B$'s sites are adjacent to one another. The other distinct partition,
where $A$ and $B$ control diagonally opposite modes never contains
entanglement in either the ground or thermal state.  At $U = 0$ we
have no entanglement of particles, as seems reasonable since the
bosons are non-interacting.

In fact it is possible to show that the ground state of the
non-interacting Bose-Hubbard model with an arbitrary number of sites
has zero entanglement of particles for any possible bi-partition.
Consider an $N$-mode ring containing $M$ non-interacting bosons. The
ground state is
\begin{equation}
\label{eq:groundfreebosons} \ket{g(N,M)} = \frac{1}{\sqrt{N^M M!}}
\left( \sum_{j=0}^{N-1} \hat{b}_j^\dagger \right)^M \ket{\rm vac},
\end{equation}
where $\ket{\rm vac}$ is the vacuum state containing zero particles in
each mode.  If $A$ controls $n_A$ modes and $B$ $n_B$ modes
($n_A+n_B=N$), and we project onto $A$ and $B$ having $m_A$ and $m_B$
bosons respectively ($m_A+m_B = M$), then using the commutation
relations amongst the boson modes
\begin{eqnarray}
\nonumber \Pi_{m_A, m_B} \ket{g(N,M)} &\propto& \left(\sum_{j \in A}
\hat{b}_j^\dagger \right)^{m_A} \left(\sum_{j \in B}
\hat{b}_j^\dagger \right)^{m_B}
\ket{\rm vac}, \\
\label{eq:factoredfreebosons} &\propto& \ket{g (n_A,m_A)} \ket{g
(n_B,m_B)}.
\end{eqnarray}
In words, the projected wavefunction is proportional to a factorized
wavefunction where $A$ or $B$'s wavefunction is the ground state of
their $m_A$ or $m_B$ non-interacting bosons as if their $n_A$ or $n_B$
modes were arranged in a ring. Therefore there is no entanglement of
particles for any division of the lattice into $A$ and $B$.  By
contrast the entanglement of modes is non-zero between any two
partitions with respect to this spatial mode decomposition, even for
non-interacting bosons.

For $U \neq 0$ we have non-zero entanglement of particles in the
ground state, which increases with $U/t$ and plateaus at approximately
$0.1405$.  The limit of large $U$ or small $t$, sometimes referred to
as the hard-core boson limit, displays interesting behaviour in terms
of entanglement of particles.  At precisely $t=0$ the ground state is
six-fold degenerate corresponding to the six ways of arranging the two
bosons in four modes such that no mode contains two bosons. Each of
these canonical ground states has zero entanglement of particles.
However there are linear superpositions of these ground states that
have non-zero $E_P$.  In particular, using degenerate perturbation
theory we find the $t \to 0$ limit of the (non-degenerate) ground
state is the superposition
\begin{eqnarray}
\nonumber \ket{g} &\to& \Big[ \ket{1,0,1,0}+\ket{0,1,0,1} - \frac{1}{\sqrt{2}} \big(\ket{1,1,0,0} \\
\label{eq:bosonground} && +\ket{0,1,1,0}+\ket{0,0,1,1}+\ket{1,0,0,1}
\big) \Big]/2.
\end{eqnarray}
The different coefficients in this sum can be understood as due to
suppressed ability to tunnel when the particles are in adjacent modes;
because $U$ is much larger than $t$ tunneling such that two particles
end up on the same site is energetically unfavorable.  From this
expression we can see why the ground state has zero entanglement of
particles in the diagonal partition ($A = \{1,3 \}$, $B = \{2,4 \}$)
but non-zero $E_P$ in the adjacent partition ($A = \{1,2 \}$, $B =
\{3,4 \}$). The projected state for the diagonal partition is
\begin{eqnarray*}
\Pi_{1,1} \ket{g} &\propto& \ket{1,1,0,0} + \ket{0,1,1,0}+\ket{0,0,1,1}+\ket{1,0,0,1} \\
&=& (\ket{0,1}+ \ket{1,0})_A (\ket{0,1}+ \ket{1,0})_B,
\end{eqnarray*}
which is separable.  For the adjacent partition the projected state is
\begin{eqnarray*}
\Pi_{1,1} \ket{g} &=& \Big[\ket{1,0,1,0}+\ket{0,1,0,1}\\
&&+\frac{1}{\sqrt{2}} \left(\ket{0,1,1,0}+\ket{1,0,0,1} \right)
\Big]/2
\end{eqnarray*}
which is non-separable --- it has $E_F = h(1/2+\sqrt{2}/3)$, where
$h(x) = -x \log_2(x) - (1-x) \log_2(1-x)$ is the binary entropy, and
normalisation $P_{1,1} = 3/4$.  Therefore entanglement of particles is
\begin{equation}
E_P(\ket{g}) = \frac{3}{4} h(1/2+\sqrt{2}/3) \simeq 0.1405.
\end{equation}

At non-zero temperature, $T$, the canonical-ensemble thermal state is
\begin{equation}
\rho = \exp(-H/k_B T) / \mathcal{Z},
\end{equation}
where $k_B$ is Boltzman's constant.  As $T \to 0$ the thermal state
approaches Eq.~(\ref{eq:bosonground}) for small non-zero $t$, and so
should contain entanglement of particles below some temperature. In
particular, for the ground state to have the majority of the weight in
the thermal-state mixture we need $k_B T$ to be of order or less than
the energy gap to the first excited state.

There are four states in the $t=0$ ground-state manifold that remain
at energy $0$ for small $t$.  To first order in $t$, the
state~(\ref{eq:bosonground}) has energy
\begin{equation}
E_g =-2 \sqrt{2} t.
\end{equation}
Therefore the energy gap for small $t$ is of order $2 \sqrt{2} t$, and
the condition for entanglement of particles is
\begin{equation}
  \label{eq:EPscaleTt} k_B T \lesssim 2 \sqrt{2} t.
\end{equation}

\begin{figure}
\centering
\includegraphics[width=80mm]{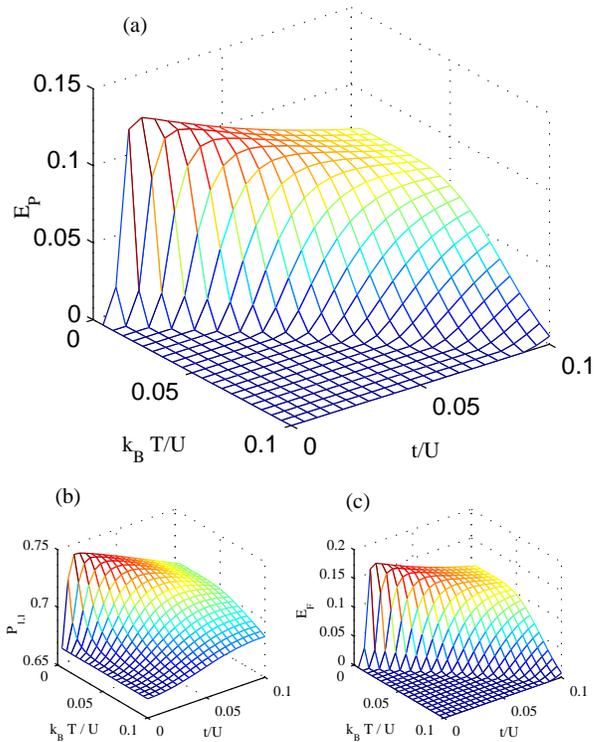}
\caption{(Color online) (a) Bipartite entanglement of particles in the
  canonical-ensemble thermal state of two bosons in four modes with
  Bose-Hubbard Hamiltonian and periodic boundary conditions.  $A$ and
  $B$ each control adjacent modes in the ring, as in
  Fig.~\ref{fig:2bosons4modesground}. (b) Probability of finding one
  particle at $A$ and one at $B$, $P_{1,1}$ . (c) Entanglement of
  formation of the \textit{a posteriori} state, $\rho^{(1,1)}_{AB}$. }
\label{fig:2bosons4modesthermal}
\end{figure}

In Fig.~\ref{fig:2bosons4modesthermal} we plot the entanglement of
particles in the thermal state as a function of the inverse
temperature and the tunneling, both scaled by $U$ to obtain
dimensionless quantities.  The plot is for small $t/U$ and we see the
type of scaling we expect --- entanglement of particles appears at
inverse temperatures proportional to the inverse of the tunneling, as
in Eq.~(\ref{eq:EPscaleTt}).

\subsection{Two spinless fermions in four modes}

We now study a precisely analogous model for fermions ---
non-interacting spinless fermions in a four-mode ring.  The
Hamiltonian is
\begin{equation}
\label{eq:freefermionfourmode} \hat{H} = -t \sum_{j=0}^3
(\hat{c}^\dagger_j \hat{c}_{j+1} + \hat{c}^\dagger_{j+1} \hat{c}_j),
\end{equation}
where $\hat{c}_j$, $\hat{c}^\dagger_j$ are fermion annihilation and
creation operators, satisfying anticommutation relations $\{
\hat{c}_j,\hat{c}_{j'}^\dagger \} = \delta_{j j'}$.  Due to the Pauli
exclusion principle it is not possible to have two spinless fermions
on the same site so there can be no on-site interaction term.

It is straightforward to calculate the spectrum by Fourier
transforming the annihilation operators, $\hat{C}_k~=~\frac{1}{2}
\sum_{j = 0}^3 e^{2 \pi i j k/ 4} \hat{c}_j, \quad k = 0 \ldots 3$.
For $N=2$ particles the ground state is two-fold degenerate --- a
basis is
\begin{equation}
\{ \ket{g_1} = C^\dagger_{1} C^\dagger_{0} \ket{{\rm
vac}}, \quad \ket{g_2} = C^\dagger_{3} C^\dagger_{0} \ket{{\rm vac}}
\}.
\end{equation}
As for bosons, the entanglement of particles in the ground or thermal
state is zero for the diagonal partition.  However for the adjacent
partition these basis states each have $P_{1,1} = 3/4$ and $E_F =
h(1/2+\sqrt{2}/3)$.  Furthermore the equal mixture of these two
states, i.e.\ the $T \to 0$ limit of the canonical-ensemble thermal
state, has the same values for $P_{1,1}$ and $E_F$~\footnote{It may
  seem unusual that a mixture of two pure states can have the same
  entanglement of formation as either of them individually; in this
  case it is because these two pure states coincidentally achieve the
  minimum in the definition of $E_F$ for the mixed
  state~\cite{wootters1998}. Using the negativity, which is an upper
  bound for the entanglement of distillation, $E_D$~\cite{vidal2002},
  it is possible to show that $E_D$ is strictly less for the mixed
  state than for either pure state.}. So we have the somewhat
surprising result that even \textit{non-interacting} fermions can have
non-zero entanglement of particles in the ground state.  This is in
stark contrast to bosons, where in the non-interacting limit the
entanglement of particles was zero.  Mathematically the reason that
non-interacting fermions can have non-zero entanglement of particles
but non-interacting bosons cannot is that the commutation relations
needed to obtain Eq.~(\ref{eq:factoredfreebosons}) as a local particle
number projection from Eq.~(\ref{eq:groundfreebosons}) do not hold for
fermions.

Motivated by this counterintuitive behaviour of non-interacting
fermions we now turn to another simple, and perhaps more
experimentally-relevant, model of two fermions in four modes --- the
Hubbard dimer.

\subsection{Hubbard Dimer}

\label{sec:hubbarddimer}

The two-site Hubbard model (Hubbard dimer) for fermions with spin
(e.g.\ electrons) is defined by the Hamiltonian
\begin{equation}
\label{eq:FHtwospinmode} \hat{H} = -t \sum_{\sigma = \up, \down}
 \left( \hat{c}^\dagger_{L \sigma} \hat{c}_{R \sigma} + \hat{c}^\dagger_{R
\sigma} \hat{c}_{L \sigma} \right) + U \sum_{j=L,R} \hat{n}_{j \up}
\hat{n}_{j \down},
\end{equation}
where $j = L, R$ is a position label and $\sigma = \up, \down$ is a
spin label.  The $t$ term describes hopping between the two sites
while conserving spin, and the $U$ term is a coloumb interaction
between fermions on the same site.

The Hubbard dimer is a simple model for a number of physical systems,
including the electrons in a $H_2$ molecule~\cite{ashcroftmermin}.  By
varying $t$ we have a model of bond breaking as the two atoms are
separated.

The ground state may be calculated exactly as a function of $U/t$, see
e.g.\ \cite{zanardi2002},
\begin{equation}
\ket{g} \propto \hat{G}_0 \ket{\rm vac},
\end{equation}
where
\begin{equation}
\label{eq:G0} \hat{G}_0 = \hat{c}^\dagger_{L \up}\hat{c}^\dagger_{L
\down} + \hat{c}^\dagger_{R \up} \hat{c}^\dagger_{R \down} + \alpha
(U/4t) ( \hat{c}^\dagger_{L \up} \hat{c}^\dagger_{R \down} -
\hat{c}^\dagger_{L \down} \hat{c}^\dagger_{R \up}),
\end{equation}
and $\alpha(x) = x + \sqrt{1 + x^2}$.

For the purposes of calculating entanglement of particles it seems
most natural to imagine party $A$ controlling the up and down modes of
one site and party $B$ the up and down modes of the other site.  With
this partition we see that the entanglement of particles in the ground
state comes entirely from the second term in~(\ref{eq:G0}), where
there is one fermion at each site forming a singlet.  The projected
state, the singlet, has entanglement of formation equal to $1$, and as
$U/t$ increases the probability $P_{1,1}$ increases from $1/2$ to $1$
as the fermions are forced to localise on each site. At a fixed
temperature we see that the entanglement of particles reaches a peak
as a function of $U/t$.  When interpreted as a model for $H_2$
bond-breaking the peak corresponds to an optimal distance at which the
trade-off between entanglement and probability of measuring one
electron at each atom is maximized.  However in reality the Hubbard
model is only a good approximation to the molecule when $U/t$ is not
too large or small~\cite{ashcroftmermin}, and at finite temperature
vibrational modes will become relevant, so it is unclear whether this
effect could actually be observed in $H_2$.

\begin{figure}
\centering
\includegraphics[width=80mm]{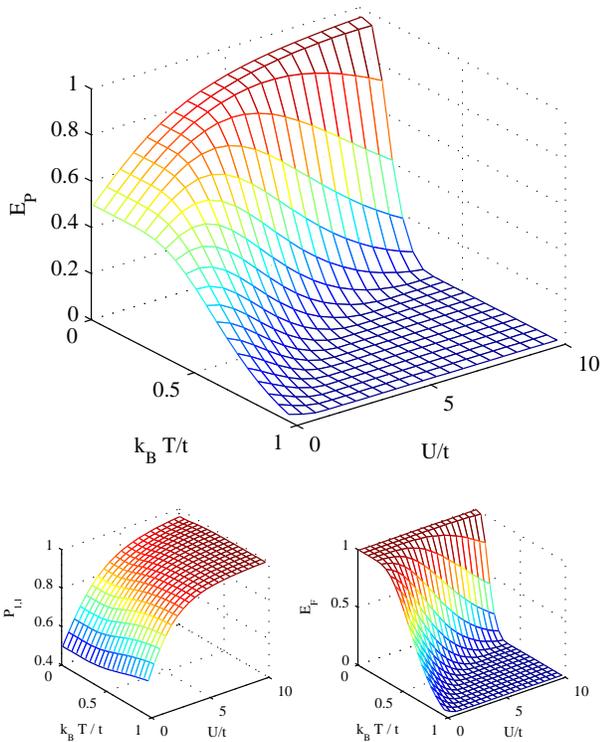}
\caption{
  (Color online) Bipartite entanglement of particles in the
  canonical-ensemble thermal state of the Hubbard dimer as a function
  of the scaled temperature, $k_B T/t$ and on-site interaction, $U/t$.
  $A$ controls the up and down modes of one site and $B$ the up and
  down modes of the other.  The color represents the probability of
  finding one particle at each site and the height represents the
  entanglement of formation of the \textit{a posteriori} state. }
\label{fig:hubbarddimer_thermal}
\end{figure}

The behaviour of the canonical-ensemble thermal state is plotted in
Fig.~\ref{fig:hubbarddimer_thermal}.  For any $U/t$ the entanglement
of formation of the projected state approaches $1$ (singlet) as the
system is cooled to the ground state ($k_B T / t \to 0$), whereas the
probability, $P_{1,1}$, approaches some value between 0 and 1 that
increases with $U/t$ (as the particles become more localized).

In \cite{zanardi2002} Zanardi performed a similar calculation of
entanglement in the ground state of the Hubbard dimer.  He calculates
what we refer to as the entanglement of modes between the two sites,
which doesn't distinguish local entropy arising from indefinite local
particle number (``charge fluctuations'') and that from entanglement
of the spins (``spin fluctuations'').  From his point of view the
ground state becomes less entangled as one increases $U/t$ as it goes
from a superposition over four local states at each site ($0$, $\up$,
$\down$, $2$) to a superposition over just two ($\up$,$\down$). From
the entanglement of particles viewpoint it is only the ``spin
fluctuations'' that are due to accessible entanglement between the two
sites, and these increase with $U/t$.

Zanardi also considers the entanglement of modes in the reciprocal
(momentum) space, where the Fourier-transformed mode operators are
$\hat{C}_{k \sigma}~=~(\hat{c}_{L \sigma}~+~e^{i k \pi} \hat{c}_{R
  \sigma} )/\sqrt{2}$, for $k=0,1$, $\sigma = \up, \down$.  In this
mode representation the operator that creates the ground state,
$\hat{G}_0$, may be written
\begin{equation}
\label{eq:G0k} \hat{G}_0 = \sum_{k = 0,1} \left[ 1+e^{i k \pi}
\alpha(U/4t) \right] \hat{C}_{k \up}^\dagger \hat{C}_{k
\down}^\dagger.
\end{equation}
We can see from this expression that the two fermions are perfectly
correlated in momentum for any $U/t$ --- both terms create the two
fermions in the same $k$ mode, one up and one down.  Therefore, if $A$
controls one $k$ mode and $B$ the other there is no entanglement of
particles in this state.

Finally, one could imagine $A$ controlling both up modes and $B$ both
down modes.  To observe this type of entanglement we could, for
example, use a magnetic field to separate up and down fermions.  We
could then look for entanglement of particles in the position or
momentum degrees of freedom, post-selected on having one up and one
down.  From Eq.~(\ref{eq:G0}) or Eq.~(\ref{eq:G0k}) we see that each
term in the superposition has one fermion up and one fermion down, so
$P_{1,1} = 1$ and the entanglement of particles coincides with the
entanglement of modes.  Since we are guaranteed to have one up and one
down, the change between position and momentum bases is a ``local''
change of basis~\footnote{Where ``local'' is defined by the spin
  variable, as that is what $A$ or $B$ each control, even though they
  are not local in space.} and the entanglement is therefore
independent of this choice.  It is equal to the mode entanglement
between the momentum modes, as calculated
in~\cite{zanardi2002}~\footnote{Zanardi uses the local entropy as his
  measure of entanglement.} --- it increases from $E_F = 0$ at $U/t =
0$ to $E_F \to 1$ as $U/t \to \infty$.

The fact that one sees entanglement in position when the particles are
distinguished by spin or entanglement in spin when the particles are
distinguished by position may be viewed as a type of ``dualism of
entanglement'', as addressed in~\cite{bose2005}.

Thus we see that there is rather subtle structure to the entanglement
in the ground state of the Hubbard dimer that is not revealed by
simply calculating the entanglement of modes.  The subtlety is above
and beyond the dependence of the mode entanglement on the choice of
modes --- it arises from considering how one might perform a
measurement in practice to reveal this entanglement and is captured by
the entanglement of particles.

\section{Multimode systems}

\label{sec:multimode}

Having studied the behaviour of the entanglement of particles in a few
small systems we now move to more-realistic systems containing many
particles in many modes.  This is typically the situation studied in
condensed matter physics; a Hamiltonian is specified in terms of
annihilation and creation operators for bosons or fermions on a
discrete set of lattice sites labeled by an index, $j$ say.  In order
to begin to get a feel for the role of entanglement in such systems
one may ask simple questions such as: is there entanglement between
the spins of fermions on two distinct lattice sites, $j_A$ and $j_B$?
If we imagine $A$ to have control of site $j_A$ and $B$ to have
control of site $j_B$ then this is precisely the situation addressed
by the entanglement of particles --- $A$ and $B$ share a state of
indistinguishable particles of indefinite local particle number.

\subsection{Entanglement of particles from correlation functions}

\label{sec:EPcorrel}

We restrict our attention to fermions for the remainder of the paper
as we have seen in Sec.~\ref{sec:smallsystems} that they can display
counterintuitive features of entanglement of particles that are not
seen for bosons. As an aid in answering questions such as the one
posed above we show how to write the projected density matrices that
appear in the definition of entanglement of particles in terms of
correlation functions.

First note there are four possible local states at each site:
$\ket{0}$, $\ket{\up}$, $\ket{\down}$ and $\ket{2}$ corresponding to
zero fermions on the site, a single fermion with spin up, a single
fermion with spin down and a doubly occupied site. Hence if one traces
out the rest of the lattice besides two sites one obtains a $16 \times
16$ reduced density matrix between those two sites.  We refer to this
matrix as the \textit{full two-site matrix} in the following.

When a local particle number superselection rule applies, the only
accessible entanglement is between the projected state with one and
only one particle at each site --- if even one of the sites contains
either zero or two fermions then there is no room for entanglement as
there is then only one possible local state at that site.  This
projected state, $\rho^{(1,1)}$ is a $4 \times 4$ matrix that we refer
to as the \textit{projected two-site matrix}. It may calculated as
\begin{equation}
\label{eq:rhositeproj} \rho^{(1,1)} = \Tr_{\overline{j_A,j_B}} [
\Pi_{1,1} \rho \Pi_{1,1} ],
\end{equation}
where $\rho$ is the total state of the system (e.g.\ in the next
subsection we will take $\rho$ to be the grand canonical ensemble
thermal density matrix), $\Tr_{\overline{j_A,j_B}}$ indicates the
trace over all sites in the lattice besides $j_A$ and $j_B$, and
$\Pi_{1,1}$ is the projector onto the subspace where there is one and
only one fermion at both $j_A$ and $j_B$.  The normalisation (trace)
of $\rho^{(1,1)}$, corresponds to the \textit{a priori} probability of
detecting one and only one particle at each site, $P_{1,1}$.

The projector, $\Pi_{1,1}$ may be written in terms of number
operators as
\begin{eqnarray}
\nonumber \Pi_{1,1} &=& \hat{n}_{j_A \up} (1-\hat{n}_{j_A \down}) \hat{n}_{j_B \up} (1-\hat{n}_{j_B \down}) \\
\nonumber && + \hat{n}_{j_A \up} (1-\hat{n}_{j_A \down}) \hat{n}_{j_B \down} (1-\hat{n}_{j_B \up}) \\
\nonumber && + \hat{n}_{j_A \down} (1-\hat{n}_{j_A \up}) \hat{n}_{j_B \up} (1-\hat{n}_{j_B \down}) \\
\label{eq:11projector} &&+ \hat{n}_{j_A \down} (1-\hat{n}_{j_A \up})
\hat{n}_{j_B \down} (1-\hat{n}_{j_B \up}).
\end{eqnarray}

The matrix elements of $\rho^{(1,1)}$ may be written as averages over
creation and annihilation operators as
 \begin{eqnarray}
\nonumber \rho^{(1,1)}_{s s', t t'} &=& \Tr_{j_A,j_B}[ \rho^{(1,1)}
\hat{c}_{j_A t}^\dagger \hat{c}_{j_B t'}^\dagger \hat{c}_{j_B s'} \hat{c}_{j_A s} ], \\
\label{eq:rhositeprojcorrel} &=& \langle \Pi_{1,1} \hat{c}_{j_A
t}^\dagger \hat{c}_{j_B t'}^\dagger \hat{c}_{j_B s'} \hat{c}_{j_A s}
\Pi_{1,1} \rangle,
\end{eqnarray}
where $s,s',t,t'$ take the values $\up,\down$. In
Table~\ref{tab:projtwositecorrel} these correlation functions are
simplified where possible by substituting (\ref{eq:11projector}) into
(\ref{eq:rhositeprojcorrel}).

\begin{table}[]
\begin{tabular}{|c|c|} \hline
site element & correlation function \\ \hline
$(\up \up, \up \up)$ & $\langle \hat{n}_{j_A \up} (1-\hat{n}_{j_A \down}) \hat{n}_{j_B \up} (1-\hat{n}_{j_B \down}) \rangle$ \\
$(\up \up, \up \down)$ & $\langle \hat{n}_{j_A \up} (1 - \hat{n}_{j_A \down}) \hat{c}^\dagger_{j_B \down}  \hat{c}_{j_B \up} \rangle$ \\
$(\up \up, \down \up)$ & $\langle \hat{c}^\dagger_{j_A \down}  \hat{c}_{j_A \up} \hat{n}_{j_B \up} (1 - \hat{n}_{j_B \down}) \rangle$ \\
$(\up \up, \down \down)$ & $\langle \hat{c}^\dagger_{j_A \up} \hat{c}_{j_A \down} \hat{c}^\dagger_{j_B \up}  \hat{c}_{j_B \down} \rangle$ \\
$(\up \down, \up \down)$ & $ \langle \hat{n}_{j_A \up} (1-\hat{n}_{j_A \down}) \hat{n}_{j_B \down} (1-\hat{n}_{j_B \up}) \rangle$ \\
$(\up \down, \down \up)$ &  $\langle \hat{c}^\dagger_{j_A \up} \hat{c}_{j_A \down} \hat{c}^\dagger_{j_B \down} \hat{c}_{j_B \up} \rangle$ \\
$(\up \down, \down \down)$ & $\langle \hat{c}^\dagger_{j_A \down} \hat{c}_{j_A \up} \hat{n}_{j_B \down} (1 - \hat{n}_{j_B \up}) \rangle$ \\
 $(\down \up, \down \up)$ & $\langle \hat{n}_{j_A \down} (1-\hat{n}_{j_A \up}) \hat{n}_{j_B \up} (1-\hat{n}_{j_B \down}) \rangle$ \\
$(\down \up, \down \down)$ & $\langle  \hat{n}_{j_A \down} (1 - \hat{n}_{j_A \up}) \hat{c}^\dagger_{j_B \down} \hat{c}_{j_B \up} \rangle$ \\
$(\down \down, \down \down)$ & $\langle \hat{n}_{j_A \down}
(1-\hat{n}_{j_A \up}) \hat{n}_{j_B \down} (1-\hat{n}_{j_B \up})
\rangle$ \\ \hline \hline
$(\up 2, \up 2)$ & $\langle \hat{n}_{j_A \up} (1 - \hat{n}_{j_A \down}) \hat{n}_{j_B \up} \hat{n}_{j_B \down} \rangle$ \\
$(\up 2, \down 2)$ & $\langle \hat{c}^\dagger_{j_A \down} \hat{c}_{j_A \up} \hat{n}_{j_B \up} \hat{n}_{j_B \down} \rangle$ \\
$(\down 2, \down 2)$ & $\langle \hat{n}_{j_A \down} (1 - \hat{n}_{j_A \up}) \hat{n}_{j_B \up} \hat{n}_{j_B \down} \rangle$ \\
$(2 \up, 2 \up)$ &  $\langle \hat{n}_{j_A \up} \hat{n}_{j_A \down} \hat{n}_{j_B \up} (1 - \hat{n}_{j_B \down}) \rangle$ \\
$(2 \up, 2 \down)$ & $\langle \hat{n}_{j_A \up} \hat{n}_{j_A \down} \hat{c}^\dagger_{j_B \up} \hat{c}_{j_B \down} \rangle$ \\
$(2 \down, 2 \down)$ & $\langle \hat{n}_{j_A \up} \hat{n}_{j_A \down} \hat{n}_{j_B \down} (1 - \hat{n}_{j_B \up}) \rangle$ \\
$(2 2, 2 2)$ & $\langle \hat{n}_{j_A \up} \hat{n}_{j_A \down}
\hat{n}_{j_B \up} \hat{n}_{j_B \down} \rangle$ \\ \hline
\end{tabular}
\caption{ Above double line: the elements of the projected two-site
matrix, $\rho^{(1,1)}$, written as correlation functions. Below
double line: elements of the full two-site matrix that contribute to
the two-spin reduced density matrix, but are not in the projected
state. The elements below the diagonal are obtained by complex
conjugation.} \label{tab:projtwositecorrel}
\end{table}

\subsection{Non-interacting electrons on a lattice}

\label{sec:freeelectrons}

Perhaps the simplest multimode fermionic system one can imagine is
non-interacting electrons in thermal equilibrium at zero or finite
temperature.  We will be interested in the thermodynamic limit ---
large lattice size --- which is of relevance to condensed matter
physics, and serves as the starting point for more realistic,
interacting, models of real materials such as superconductors.  We aim
to clarify issues of entanglement in this simple case in order that
the same issues can be addressed in interacting systems.

The Hamiltonian for non-interacting electrons is
\begin{equation}
\label{eq:Hfreefermions} \hat{H} = - t \sum_{\substack{<j,k> \\
\sigma = \up, \down}} \hat{c}_{j \sigma}^\dagger \hat{c}_{k \sigma},
\end{equation}
where $< j,k >$ indicates that the sum runs over nearest neighbors
$j$ and $k$ as defined by a link in the lattice.  If we were to add an
on-site interaction between electrons of opposite spin (i.e.\ a
coloumb interaction), $U \sum_{j} \hat{n}_{j \up} \hat{n}_{j \down}$,
we would have the well-studied Hubbard model, the two-site version of
which was studied in Sec.~\ref{sec:hubbarddimer}.  This simple type of
interaction might be a good starting point for studying how
interactions affect the entanglement of particles.

One's immediate reaction may be that there can be no entanglement in
this system as the fermions are non-interacting and the up and down
spins are independent.  However this intuition was shown to be
incorrect in Sec.~\ref{sec:hubbarddimer}, where we saw that the
two-site version of this model ($U = 0$) contained entanglement of
particles in the ground state.  Furthermore, in \cite{vedral2003,
  oh2004} it is argued that it is indeed possible to have
``entanglement of spins in a non-interacting electron gas'', which is
roughly the continuum limit of our discrete model. There are subtle
differences between the entanglement as studied in those works and the
concept of entanglement of particles that we have focused on here.
Sec.~\ref{sec:continuum} contains a detailed comparison of this
previous work to the current entanglement-of-particles approach.

We now explicitly calculate the entanglement of particles for
non-interacting fermions on a lattice, Eq.~(\ref{eq:Hfreefermions}),
in thermal and chemical equilibrium.  We choose the simple case of a
1-D lattice of $M$ sites with closed boundary conditions (i.e.\ a
ring) for the purpose of illustration.  The state of the system at
temperature $T$ and chemical potential $\mu$ is given by the grand
canonical ensemble density matrix
\begin{equation}
\label{eq:gcedensitymatrix} \rho_T = \exp(-(\hat{H} - \mu \hat{N})/
k_B T)/\mathcal{Z},
\end{equation}
where $\hat{N}$ is the total number operator and $\mathcal{Z}$ is the
grand canonical partition function,
$\mathcal{Z}~=~\Tr[\exp(-(\hat{H}~-~\mu \hat{N})/k_B T)]$.

In order to explicitly calculate $\rho^{(1,1)}$ we use the fact that
the grand canonical ensemble density matrix,
Eq.~(\ref{eq:gcedensitymatrix}), is a Gaussian state when the system
is described by the non-interacting fermion Hamiltonian,
Eq.~(\ref{eq:Hfreefermions}). For this reason higher-order correlation
functions, as in Table~\ref{tab:projtwositecorrel} factorize into
second-order correlation functions.

Another simplifying feature is that many of the matrix elements are
zero due to the collective $SU(2)$ rotational symmetry of the model.
In fact the only non-zero elements are those along the diagonal, and
the off-diagonal elements $\rho^{(1,1)}_{\up \down, \down \up} =
\rho^{(1,1) *}_{\down \up, \up \down}$.  A way to see this is via the
well-known result due to Weyl that states invariant under collective
$SU(2)$ rotations of the spin have the form
\begin{equation}
\label{eq:werner} \rho = \int dU U \otimes U \rho U^\dagger \otimes
U^\dagger = p_A \Pi_A + p_S \Pi_S
\end{equation}
where $\Pi_{A/S}$ are the projectors onto the antisymmetric (spanned
by the singlet) and symmetric (spanned by the three triplet states)
subspaces, and $p_{A/S}$ are the weights of these projectors.  In our
case $p_A + p_S = P_{1,1}$.  States of this form are known as
\textit{Werner states} in quantum information theory.

The non-zero matrix elements are all determined by two second order
correlation functions
\begin{eqnarray}
\bar{n} &=& \langle \hat{n}_{j_A \up} \rangle = \langle \hat{n}_{j_A \down} \rangle = \langle \hat{n}_{j_B \up} \rangle = \langle \hat{n}_{j_B \down} \rangle, \\
c_{j_A,j_B} &=& \langle \hat{c}^\dagger_{j_A \up} \hat{c}_{j_B \up}
\rangle = \langle \hat{c}^\dagger_{j_A \down} \hat{c}_{j_B \down}
\rangle,
\end{eqnarray}
where for the first line we have also used the translational
invariance of the lattice.  The average occupation of any individual
up or down mode which we call the \textit{filling factor} in the
following, is given by $\bar{n}$, and $c_{j_A,j_B}$ is an exchange
correlation between the two sites.  Explicitly the matrix elements are
\begin{eqnarray}
\nonumber \rho^{(1,1)}_{\up \up, \up \up} &=& \rho^{(1,1)}_{\down \down, \down \down} = (\bar{n}^2 - |c_{j_A,j_B}|^2)( (1-\bar{n})^2  - |c_{j_A,j_B}|^2) \\
\nonumber \rho^{(1,1)}_{\up \down, \up \down} &=& \rho^{(1,1)}_{\down \up, \down \up} = ( \bar{n} (1 - \bar{n}) + |c_{j_A,j_B}|^2 )^2 \\
\label{eq:matrixelements} \rho^{(1,1)}_{\up \down, \down \up} &=& -
|c_{j_A,j_B}|^2.
\end{eqnarray}

For non-interacting fermions we can actually calculate the two
correlation functions, $\bar{n}$ and $c_{j_A,j_B}$, explicitly as a
function of $\mu$ and $T$.  The Hamiltonian (\ref{eq:Hfreefermions})
is diagonal when written in terms of momentum creation and
annihilation operators
\begin{equation}
\hat{H} = -2 t \sum_{\substack{k = 0 \\ \sigma = \up, \down}}^{M-1}
\cos(2 \pi k / M) \hat{C}^\dagger_{k \sigma} \hat{C}_{k \sigma},
\end{equation}
where
\begin{equation}
\hat{C}_{k, \sigma} = \frac{1}{\sqrt{M}} \sum_{j = 0}^{M-1} e^{2 \pi
i j k/ M} \hat{c}_{j \sigma}.
\end{equation}
The occupation of the momentum-space modes is therefore
\begin{equation}
n_k = \langle \hat{C}^\dagger_{k \up} \hat{C}_{k \up} \rangle =
\langle \hat{C}^\dagger_{k \down} \hat{C}_{k \down} \rangle =
\frac{1}{e^{-(E_k - \mu)/ k_B T} + 1}
\end{equation}
where $E_k = -2 t \cos(2 \pi k / M)$ is the energy of the $k^{\rm th}$
mode.  By inverse Fourier transforming back to the position-space
modes we obtain
\begin{eqnarray}
\bar{n} &=& \frac{1}{M} \sum_{k = 0}^{M-1} n_k \\
c_{j_A,j_B} &=& \frac{1}{M} \sum_{k = 0}^{M-1} e^{2 \pi i (j_A -j_B) k} n_k.
\end{eqnarray}

\begin{figure}
\centering
\includegraphics[width=80mm]{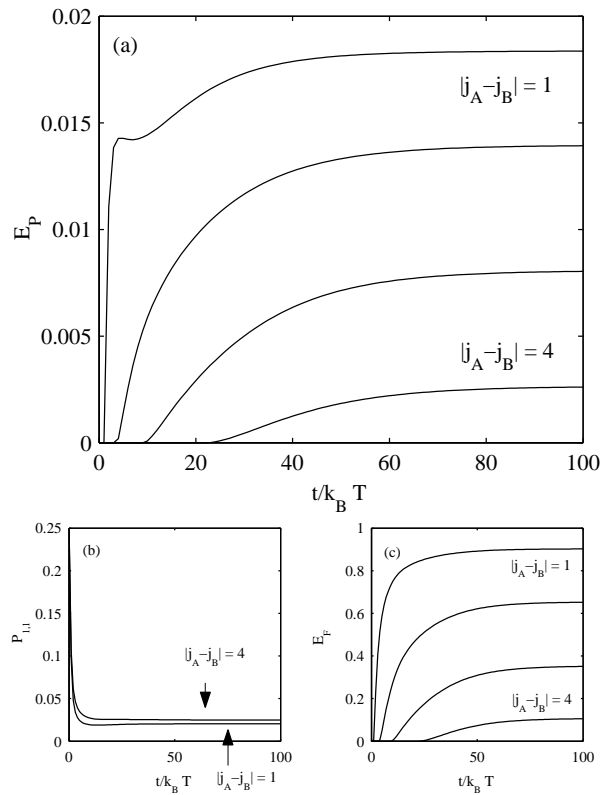}
\caption{(a) Entanglement of particles for non-interacting fermions
  on a $1-D$ lattice with $30$ sites as a function of inverse
  temperature ($t/k_B T$) and separation between the sites ($|j_A -
  j_B|$).  The subfigures (b) and (c) indicate the contributions to
  $E_P$ from the probability of finding one particle at each site,
  $P_{1,1}$, and the entanglement of formation in the \textit{a
    posteriori} state, respectively.  The chemical potential was $\mu
  \simeq -1.89$ corresponding to a $T \to 0$ filling factor of
  $\bar{n} = 0.2$. In the $T \to 0$ limit we therefore roughly expect
  that $P_{1,1} \sim (0.2)^2 = 0.04$} \label{fig:FHN30cool}
\end{figure}

Fig.~\ref{fig:FHN30cool} (a) illustrates entanglement of particles
between two sites as the system is cooled from high temperature down
to zero temperature (the ground state) as a function of the inverse
temperature $t / k_B T$, for a fixed chemical potential (which
determines the $T \to 0$ filling factor).  At high temperature ($t /
k_B T = 0$) the probability of each up or down mode being occupied is
$0.5$ and completely uncorrelated with any other mode, so $P_{1,1} =
0.25$ and there is no entanglement. As we cool to lower temperatures
entanglement appears between increasingly distant sites in the
lattice, but the probability $P_{1,1}$ decreases because the mean atom
number is decreasing.  The entanglement in $\rho^{(1,1)}$ decreases
with the separation between the two sites and goes to zero at some
finite separation between the sites that depends on the temperature.
We call this distance the \textit{entanglement length}, $r_e$, in
anticipation of a relationship to previous work on the free Fermi gas
to be discussed in the next section.  In this case the entanglement
length is $5$ sites in the $T \to 0$ limit (ground state).

\begin{figure}
\centering
\includegraphics[width=80mm]{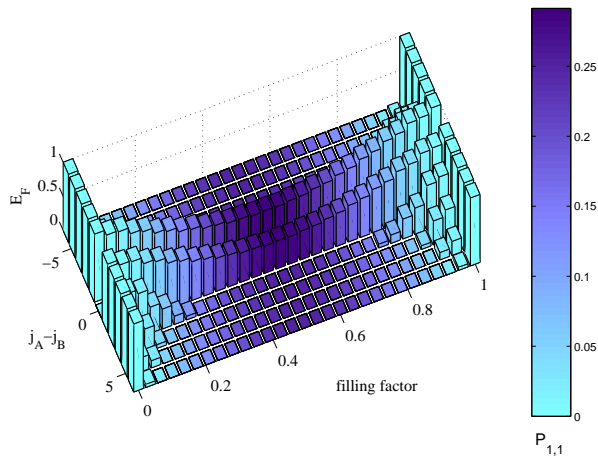}
\caption{ (Color online) Bipartite entanglement of formation
  (height) and probability, $P_{1,1}$ (greyscale) as a function of
  separation and filling factor in the $T \to 0$ limit of the grand
  canonical-ensemble thermal state of the Fermi-Hubbard model with
  $30$ sites and periodic boundary conditions.  The chemical potential
  was varied from $-2$ to $2$ to achieve different filling factors.}
\label{fig:FHN30gsentff}
\end{figure}

Fig.~\ref{fig:FHN30gsentff} illustrates the entanglement of particles
between sites in the ground state as a function of the filling factor
$\bar{n}$.  For filling factors less than one half the entanglement
length decreases with increasing filling factor, and by the time half
filling is reached there is only entanglement between neighboring
sites.  For filling factors greater than one half the entanglement
length increases again due to the particle-hole symmetry of the model
--- the filling factor for holes is decreasing.  By contrast the
probability that one particle will be found at each site reaches a
maximum at half-filling, as indicated by greyscale in the figure.

When there are only two electrons or holes in the lattice ($\bar{n} =
2/(2\times 30) = 1/30$) they form a singlet with $E_F = 1$ independent
of the separation between the sites --- i.e.\ the entanglement length
is infinite.  This effect is rather more subtle than simply
entanglement between sites as may be seen in spin models.  The
wavefunction is such that the fermions are equally likely to be found
anywhere in the lattice (apart from on top of each other which is
slightly more likely), but wherever they are found they must be in a
singlet.

Note that from the entanglement of particles perspective entanglement
is only possible if $A$ and $B$ each control both up and down modes of
\textit{distinct} sites.  If $A$ controlled the up mode and $B$ the
down mode of the \textit{same} site they could never share any
entanglement as they each only have one mode. This is why $|j_A - j_B|
= 0$ is not plotted in these figures.

\subsection{The continuum limit}

\label{sec:continuum}

We now discuss the continuum limit of the non-interacting fermion
lattice model and contrast the entanglement of particles with
another approach that has been used in recent work
--- the so-called two-spin reduced density matrix~\cite{vedral2003,
oh2004}.

The \textit{two-spin reduced density matrix} between two points
$\vec{r}$ and $\vec{r}'$ is defined as
\begin{equation}
\label{eq:rhospincont} \rho^{\rm spin}_{s s', t t'} = \langle
\hat{\psi}_t^\dagger (\vec{r}) \hat{\psi}_{t'}^\dagger(\vec{r}')
\hat{\psi}_{s'} (\vec{r}') \hat{\psi}_s (\vec{r}) \rangle,
\end{equation}
where $\hat{\psi}_s(\vec{r}), \hat{\psi}_s^\dagger(\vec{r})$ are
field annihilation/creation operators for a particle with spin $s$
located at position $\vec{r}$ satisfying $\{\hat{\psi}_s (\vec{r}),
\hat{\psi}_{s'}^\dagger(\vec{r}') \} = \delta_{s s'}
\delta(\vec{r}-\vec{r}')$.  We refer to this matrix as the
\textit{spin-correlation matrix} in what follows as we believe the
name is more appropriate.

It was shown in~\cite{vedral2003, oh2004} that for a free Fermi gas in
thermal equilibrium the spin-correlation matrix takes the form of a
Werner state, for the same reasons as for the lattice model in the
previous section.  After normalisation, let $p=p_A$ be the weight of
the singlet.  The function $p$ depends on the relative distance $r =
|\vec{r} - \vec{r}'|$ and temperature, $T$; $p=1$ at $r=0$ and $p \to
0$ as $r \to \infty$.  The \textit{entanglement length}, $r_e$, is
uniquely determined by
\begin{equation}
p(r_e,T) = 1/3,
\end{equation}
and the spins are entangled for $r<r_e$ and separable for $r \geq
r_e$.  At zero temperature the relevant parameter is the Fermi
momentum $k_F$ and the entanglement length scales as
\begin{equation}
r_e \propto 1/k_F.
\end{equation}

One might expect to see similar behaviour in a lattice model of
non-interacting fermions, and indeed we saw in the previous section
that the concept of an entanglement length persists when one considers
entanglement of particles on a lattice.  We will see subsequently that
the lattice filling factor, $\bar{n}$, plays the role of the Fermi
momentum in a certain limit.  However first we discuss the precise
relationship between the spin correlation matrix and the projected
two-site matrix.

In analogy to Eq.~(\ref{eq:rhospincont}) one may be tempted to write
down
\begin{equation}
\label{eq:rhospin} \rho^{\rm spin}_{s s', t t'} = \langle
\hat{c}_{j_A t}^\dagger \hat{c}_{j_B t'}^\dagger \hat{c}_{j_B s'}
\hat{c}_{j_A s} \rangle,
\end{equation}
as a ``two-spin reduced density matrix'' in the lattice.  However from
the results of the previous section this matrix does not not
correspond to the density matrix of a two-component quantum system in
the usual sense of quantum information theory --- the central reason
being that the two subsystems are not well-defined. By contrast the
entanglement in Eq.~(\ref{eq:rhositeprojcorrel}) is experimentally
accessible.

To see this point first note that~(\ref{eq:rhositeprojcorrel}) has the
same form as~(\ref{eq:rhospin}) but with projectors onto the
one-particle subspace inserted.  As we have argued in
Sec.~\ref{sec:review}, the projector is necessary in order to define
the subsystems; without it multi-particle correlations contribute.  If
one were to imagine extracting entangled fermions from the lattice
then the very act of extracting is implicitly a measurement of local
particle number, and entanglement can then exist only if precisely one
fermion is found per site.

Of course the correlation functions that make up the matrix could, in
principle, be measured without actually extracting
fermions~\footnote{For example by using detectors sensitive to the
  presence of either up spins or down fermions independent of whether
  the other is present.}.  Moreover, it is not difficult to show that
entanglement in the matrix~(\ref{eq:rhospin})(normalized, and treated
as if it were a density matrix) provides a lower bound on the
entanglement in the projected two-site matrix~\footnote{P.~Samuelsson
  and M.~B\"uttiker, private communication.}  Therefore if the
matrix~(\ref{eq:rhospin}) were reconstructed experimentally, as may be
possible in the near future in mesoscopic
systems~\cite{samuelsson2006}, then a calculation of non-zero
entanglement in this matrix would imply that the accessible
entanglement would be non-zero (but the converse is not true).
Nevertheless, it would, in our opinion, still be wrong to call the
spin correlation matrix~(\ref{eq:rhospin}) a density matrix, for the
reasons given above.

There are some more intuitive reasons why we should not expect
correlations between two fermions on one site and one or two fermions
on the other to contribute to entanglement.  The spatial wavefunctions
of two fermions on the same site are identical and therefore their
spin wavefunction is a singlet (as in \cite{vedral2003, oh2004}).  By
the monogamy of entanglement neither can be entangled in spin with any
other.  In this sense a doubly occupied site is like an unoccupied
site --- one should not include correlations from it in the
calculation of a density matrix. The projected two-site matrix
respects particle-hole symmetry in this sense and therefore fits in
naturally with experimental considerations in mesoscopic
systems~\cite{beenakker2005}, whereas the spin-correlation matrix does
not.

In Table~\ref{tab:elementmap} we show precisely the relationship
between the two matrices by writing the elements of the spin
correlation matrix as sums of elements of the full two-site matrix.
This mapping is not a ``coarse graining'' in the sense that each
element of the full two-site matrix contributes to only one element in
the spin-correlation matrix --- certain elements, for example $(2 2, 2
2)$, map to many of the spin elements.  Therefore the spin-correlation
matrix is not the ``density matrix'' of a well defined two-component
system in the sense that is used in quantum information.  We conclude
that, at least in lattice models, the entanglement of particles ---
i.e.\ the entanglement in the projected two-site matrix, is what
should be used instead of the entanglement in the spin correlation
matrix.

\begin{table}[]
\begin{tabular}{|c|c|} \hline
spin element & sum of site elements \\ \hline
(\up \up, \up \up)& (\up \up,\up \up) + (\up 2,\up 2) + (2 \up,2 \up) + (2 2,2 2) \\
(\up \up,\up \down)& (\up \up,\up \down) + (2 \up,2 \down)  \\
(\up \up,\down \up)& (\up \up,\down \up) + (\up 2,\down 2)  \\
(\up \up,\down \down)& (\up \up,\down \down) \\
(\up \down,\up \down)& (\up \down,\up \down) + (\up 2,\up 2) + (2 \down,2 \down) + (2 2,2 2) \\
(\up \down,\down \up)& (\up \down, \down \up)  \\
(\up \down,\down \down)&  (\up \down,\down \down) + (\up 2,\down 2) \\
(\down \up,\down \up)& (\down \up,\down \up) + (\down 2,\down 2) + (2 \up,2 \up) + (2 2,2 2) \\
(\down \up,\down \down)&  (\down \up,\down \down) + (2 \up,2 \down)  \\
(\down \down,\down \down)&  (\down \down,\down \down) + (\down
2,\down 2) + (2 \down,2 \down) + (2 2,2 2) \\ \hline
\end{tabular}
\caption{ Mapping the full two-site matrix onto the spin-correlation
matrix.  The left column represents the elements of the
spin-correlation matrix, which are obtained by summing elements of
the full two-site matrix represented in the right hand column.
} \label{tab:elementmap}
\end{table}

Despite these problems with using the spin-correlation matrix to
calculate entanglement for discrete lattice models, in the continuum
limit Eq.~(\ref{eq:rhospincont}) recovers a interpretation as a
density matrix of two spins.  The reason for this is basically that
the probability of finding two electrons at a particular location in
space is negligible compared with the probability of finding one
electron.

To see this in more detail, let $N$ non-interacting electrons be
confined to a region $[0,L]$ in one dimension.  Define a set of $M$
orthonormal wavefunctions, $\{\psi_j(x), \quad j = 0 \ldots M-1 \}$,
on the region by
\begin{eqnarray}
  \psi_j(x)  &=&  \left\{ \begin{array}{l}
      1/\sqrt{\epsilon}, \quad  x \in [j \epsilon, (j+1) \epsilon] \\
      0, \quad x \notin [j \epsilon, (j+1) \epsilon]
      \end{array}
      \right.
\end{eqnarray}
where $\epsilon = L/M$.  Let $\hat{c}_{j \sigma}$ be the annihilation
operator that destroys an electron in the $j$th region with spin
$\sigma$. For non-interacting electrons in thermal equilibrium the
probability of finding one electron in the $j$th region is $\langle
\hat{n}_{j \sigma} \rangle = N \epsilon / L$, while the probability of
finding two electrons, one up and one down, scales as $\langle
\hat{n}_{j \up} \hat{n}_{j \down} \rangle = O(\epsilon^2)$.  Therefore
in the $\epsilon \to 0$ limit, where $\psi_j(x)$ approaches a delta
function at $x = j \epsilon $, the probability of finding two
electrons at the same site becomes negligible compared to the
probability of finding one.  Hence the projector in
Eq.~(\ref{eq:rhositeprojcorrel}) has no effect, and the projected
two-site matrix approaches the spin-correlation
matrix~(\ref{eq:rhospincont}) in the continuum limit.

The question of using~(\ref{eq:rhospincont}) as a density matrix was
considered in the appendix of~\cite{cavalcanti2005}.  Their
explanation agrees with ours for the case where there is one and only
one particle in each of local modes.  Our results show why this is a
valid assumption in the continuum limit, but also apply to more
general situations when there is a non-zero probability of finding two
electrons in the same local mode.

\begin{figure}
\centering
\includegraphics[]{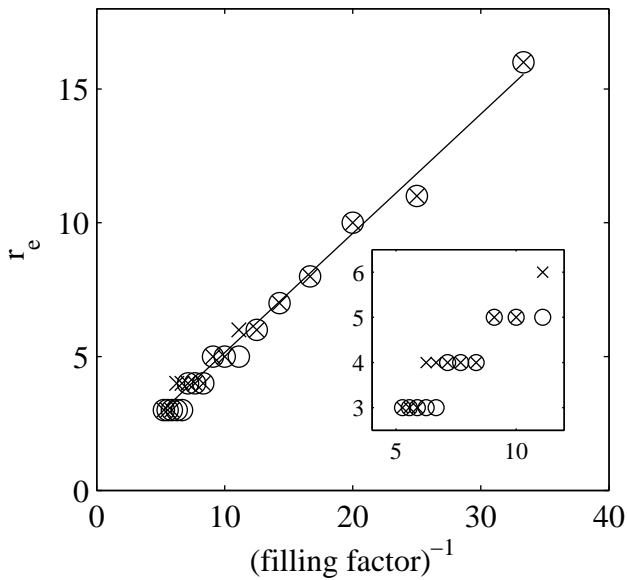}
\caption{Entanglement length (crosses), $r_e$, versus inverse
  filling factor, $1/\bar{n}$, for non-interacting fermions on a $1-D$
  lattice.  The circles show what the entanglement length would have
  been if it were defined using the spin-correlation matrix,
  Eq.~(\ref{eq:rhospin}), instead of the projected matrix,
  Eq.~(\ref{eq:rhositeprojcorrel}). The line is a linear least-squares
  fit to the crosses. The inset shows that the projected matrix
  predicts a longer entanglement length for some values of the filling
  factor.} \label{fig:re}
\end{figure}

With this calculation in mind we should expect that in the limit of
low filling factor on the lattice the difference between
Eq.~(\ref{eq:rhositeprojcorrel}) and Eq.~(\ref{eq:rhospin}) will be
negligible.  We define the entanglement length for the lattice as the
smallest $|j_A-j_B|$ for which the entanglement of the two-site matrix
is zero.  Fig.~\ref{fig:re} shows the entanglement length versus the
inverse of the filling factor alongside the ``entanglement'' length
calculated from the spin correlation matrix.  We see that in the limit
of small filling factor ($1/\bar{n}$ large) the entanglement length
scales linearly with $1/\bar{n}$, as one might expect from
\cite{vedral2003,oh2004} since $k_F \propto \bar{n}$ in one dimension.
 
For non-interacting electrons we see from Tab.~\ref{tab:elementmap}
that the many-electron correlations simply add a term proportional to
the identity to the density matrix.  This can only have the effect of
diluting the entanglement and decreasing the entanglement length. We
see from the inset in the figure that the projected matrix therefore
predicts a longer entanglement length than for the spin-correlation
matrix for some values of the filling factor approaching half-filling
($\bar{n} = 1/2$).  For $0.45 \leq \bar{n} \leq 0.55$ (not plotted in
the figure) the spin-correlation matrix would have predicted an
entanglement length of $1$ site (i.e.\ not even nearest-neighbors are
entangled), whereas the projected matrix gives $r_e=2$
(nearest-neighbors are entangled).  This is precisely the limit where
there is a good chance of finding two electrons at the same site.

\section{Summary and conclusion}

An idea that has attracted much attention recently from the quantum
information theory community is that the improved understanding of
entanglement that has been developed may lead to new insights into the
physics of strongly-correlated systems in condensed matter.  Typically
in these systems the particles are itinerant and so must be treated as
indistinguishable. Thus if one wishes to explore the role of
entanglement one must have a good understanding of what it means for
indistinguishable particles to be entangled.

We have argued that in many situations of interest in condensed matter
a local particle number superselection rule applies to operations,
such as measurements, that can be performed on the system. Thus the
notion of entanglement of particles, introduced in~\cite{wiseman2003},
may be a more appropriate measure of entanglement to use in studying
these systems than the entanglement of modes which appears in many
previous studies.  The physical meaning of entanglement of particles
is clear --- the subsystems that are supposed to be entangled are
established by measurement of local particle number by the two
parties.  The mathematical entanglement in occupation number (mode
entanglement) may still display interesting behaviour, however its
physical significance is less clear as the measurements that could be
performed to observe it are unspecified.

In order to get a feel for how the entanglement of particles compares
with the entanglement of modes we began by studying some simple
systems that are analytically solvable.  The minimal situation in
which entanglement of particles is possible is two particles in four
modes.  We found that the entanglement of particles was zero for
bosons but non-zero for fermions for two non-interacting particles in
a four-mode ring.  In both cases the mode entanglement was non-zero.
For the Hubbard dimer we calculated the entanglement of particles
according to a number of different mode decompositions and compared
with previous studies of mode entanglement~\cite{zanardi2002}.

Finally we studied non-interacting fermions on a lattice and compared
with previous results for this system.  We first showed how to write
the projected matrix for one fermion on each of two distinct sites in
terms of correlation functions.  In agreement with previous results
regarding the free Fermi gas~\cite{vedral2003,oh2004} we found an
``entanglement length'' in the system beyond which the fermions are
not entangled.  It is intriguing that this length extends over
multiple lattice sites and persists even when one considers the more
restrictive criteria of entanglement of particles where real
measurements are considered.  Thus the phenomena of entanglement of
non-interacting fermions should be experimentally observable, perhaps
in optical lattice set-ups where condensed matter Hamiltonians may be
engineered.  It cannot be dismissed as trivially due to the
antisymmeterization of the wavefunction and unobservable.  Finally we
showed precisely how the entanglement of particles relates to the
two-spin spin reduced density matrix~\cite{vedral2003,oh2004} in the
continuum limit.

Recently there has been interest in studying scaling laws for
entanglement entropy in arbitrary
dimensions~\cite{plenio2005,wolf2005}.  The entanglement entropy
corresponds to the entanglement of modes for pure states when
entanglement of formation is used as the measure.  An interesting
direction for future research would be to see if such scaling laws
persist when the more-restrictive criteria of entanglement of
particles is used, rather than entanglement of modes.

In conclusion, we believe that the entanglement of particles may be a
useful concept to consider alongside mode entanglement in studying
systems of indistinguishable particles that are central to condensed
matter physics. We have showed that it is physically well-motivated by
measurement considerations and leads to distinct phenomenology of
entanglement in a few simple systems.

\begin{acknowledgements}
  We thank Steve Bartlett, \v Caslav Brukner, Carlo Beenakker, Markus
  B\"uttiker, Daniel Cavalcanti, Joel Corney, Tom Kaplan, Damien Pope,
  Peter Samuelsson, and Vlatko Vedral for valuable discussions.
\end{acknowledgements}

\appendix


\begin{thebibliography}{99}

\expandafter\ifx\csname natexlab\endcsname\relax\def\natexlab#1{#1}\fi
\expandafter\ifx\csname bibnamefont\endcsname\relax
  \def\bibnamefont#1{#1}\fi
\expandafter\ifx\csname bibfnamefont\endcsname\relax
  \def\bibfnamefont#1{#1}\fi
\expandafter\ifx\csname citenamefont\endcsname\relax
  \def\citenamefont#1{#1}\fi
\expandafter\ifx\csname url\endcsname\relax
  \def\url#1{\texttt{#1}}\fi
\expandafter\ifx\csname urlprefix\endcsname\relax\def\urlprefix{URL }\fi
\providecommand{\bibinfo}[2]{#2}
\providecommand{\eprint}[2][]{\url{#2}}

\bibitem[{\citenamefont{Osborne and Nielsen}(2002)}]{osborne2002}
\bibinfo{author}{\bibfnamefont{T.~J.} \bibnamefont{Osborne}} \bibnamefont{and}
  \bibinfo{author}{\bibfnamefont{M.~A.} \bibnamefont{Nielsen}},
  \bibinfo{journal}{Phys. Rev. A} \textbf{\bibinfo{volume}{66}},
  \bibinfo{pages}{032110} (\bibinfo{year}{2002}).

\bibitem[{\citenamefont{Osterloh et~al.}(2002)\citenamefont{Osterloh, Amico,
  Falci, and Fazio}}]{osterloh2002}
\bibinfo{author}{\bibfnamefont{A.}~\bibnamefont{Osterloh}},
  \bibinfo{author}{\bibfnamefont{L.}~\bibnamefont{Amico}},
  \bibinfo{author}{\bibfnamefont{G.}~\bibnamefont{Falci}}, \bibnamefont{and}
  \bibinfo{author}{\bibfnamefont{R.}~\bibnamefont{Fazio}},
  \bibinfo{journal}{Nature} \textbf{\bibinfo{volume}{416}},
  \bibinfo{pages}{608} (\bibinfo{year}{2002}).

\bibitem[{\citenamefont{Vidal et~al.}(2003)\citenamefont{Vidal, Latorre, Rico,
  and Kitaev}}]{vidal2003}
\bibinfo{author}{\bibfnamefont{G.}~\bibnamefont{Vidal}},
  \bibinfo{author}{\bibfnamefont{J.~I.} \bibnamefont{Latorre}},
  \bibinfo{author}{\bibfnamefont{E.}~\bibnamefont{Rico}}, \bibnamefont{and}
  \bibinfo{author}{\bibfnamefont{A.}~\bibnamefont{Kitaev}},
  \bibinfo{journal}{Phys. Rev. Lett.} \textbf{\bibinfo{volume}{90}},
  \bibinfo{pages}{227902} (\bibinfo{year}{2003}).

\bibitem[{\citenamefont{Latorre et~al.}(2004)\citenamefont{Latorre, Rico, and
  Vidal}}]{latorre2003a}
\bibinfo{author}{\bibfnamefont{J.~I.} \bibnamefont{Latorre}},
  \bibinfo{author}{\bibfnamefont{E.}~\bibnamefont{Rico}}, \bibnamefont{and}
  \bibinfo{author}{\bibfnamefont{G.}~\bibnamefont{Vidal}},
  \bibinfo{journal}{Quant. Inf. Comp.} \textbf{\bibinfo{volume}{4}},
  \bibinfo{pages}{048} (\bibinfo{year}{2004}).

  
\bibitem{verstraete2004} F.~Verstraete, M.~A.~Martin-Delgado,
  J.~I.~Cirac, \prl {\bf 92}, 087201 (2004).
  
\bibitem{popp2005} M.~Popp, F.~Verstraete, M.~A.~Martin-Delgado,
  J.~I.~Cirac, \pra {\bf 71}, 042306 (2005).

\bibitem{tan1991} S.~M.~Tan, D.~F.~Walls, and M.~J.~Collett, \prl {\bf
    66}, 252 (1991).

\bibitem{hardy1994} L. Hardy, \prl {\bf 73}, 2279 (1994).

\bibitem{greenberger1995} D.M. Greenberger, M. A. Horne, and A.
  Zeilinger, \prl {\bf 75}, 2064 (1995).

\bibitem{vanenk2005} S.~J.~van~Enk, \pra {\bf 72}, 064306 (2005).
  
\bibitem{paskauskas2001} R.~Pa\v skauskas and L.~You, \pra {\bf 64},
  042310 (2001).
  
\bibitem{schliemann2001} J.~Schliemann, J.~I.~Cirac, M.~Kus,
  M.~Lewenstein, and D.~Loss, \pra {\bf 64}, 022303 (2001).
  
\bibitem{li2001} Y.~S.~Li, B.~Zeng, X.~S.~Liu, and G.~L.~Long, \pra
  {\bf 64}, 054302 (2001).
  
\bibitem{gittings2002} J.~R.~Gittings and A.~J.~Fisher, \pra {\bf 66},
  032305 (2002).
  
\bibitem{vedral2003} V.~Vedral, CEJP {\bf 2}, 289-306 (2003).
  
\bibitem{shi2003} Yu~Shi, \pra {\bf 67}, 024301 (2003).
  
\bibitem{wiseman2003} H.~M.~Wiseman and John~A.~Vaccaro, \prl {\bf
    91}, 097902 (2003).
  
\bibitem{kaplan2005} T.~A.~Kaplan, Fluctuation and Noise Letters {\bf
    5}, C15 (2005).
  
\bibitem{samuelsson2003} P.~Samuelsson, E.~V.~Sukhorukov, M.~Buttiker,
  \prl {\bf 91}, 157002 (2003).

\bibitem{beenakker2003} C.~W.~J.~Beenakker, C.~Emary, M.~Kindermann,
  and J.~L.~van~Velsen, \prl {\bf 91}, 147901 (2003)

\bibitem{samuelsson2004} P.~Samuelsson, E.~V.~Sukhorukov, M.~Buttiker,
  \prl {\bf 92}, 026805 (2004)

\bibitem{beenakker2005} C.~W.~J.~Beenakker, e-print cond-mat/0508488.
  
\bibitem{samuelsson2005} P.~Samuelsson, E.~V.~Sukhorukov, M.~Buttiker,
  New.~J.~Phys. {\bf 7}, 176 (2005).

\bibitem{samuelsson2006} P.~Samuelsson and M.~Buttiker, \prb {\bf 73},
  041305(R) (2006)

\bibitem{oh2004} Sangchul~Oh and Jaewan~Kim, \pra {\bf 69}, 054305
  (2004).

\bibitem{zanardi2002} Paolo~Zanardi, \pra {\bf 65}, 042101(R) (2002).
  
\bibitem[{\citenamefont{Hines et~al.}(2003)\citenamefont{Hines,
      McKenzie, and Milburn}}]{hines2003a}
  \bibinfo{author}{\bibfnamefont{A.~P.} \bibnamefont{Hines}},
  \bibinfo{author}{\bibfnamefont{R.~H.} \bibnamefont{McKenzie}},
  \bibnamefont{and} \bibinfo{author}{\bibfnamefont{G.~J.}
    \bibnamefont{Milburn}}, \bibinfo{journal}{Phys. Rev. A}
  \textbf{\bibinfo{volume}{67}}, \bibinfo{pages}{013609}
  (\bibinfo{year}{2003}).

\bibitem{zanardi2001} P.~Zanardi, \prl {\bf 87}, 077901 (2001).

\bibitem{vanenk2003} S.~J.~van~Enk, \pra {\bf 67}, 022303 (2003).
  
\bibitem{bartlett2003} Stephen~D.~Bartlett and H.~M.~Wiseman, \prl
  {\bf 91}, 097903 (2003).
  
\bibitem{aharonov1967} Y.~Aharonov and L.~Susskind, Phys.~Rev. {\bf
    155}, 1428 (1967).
  
\bibitem{bennett1996} C.~H.~Bennett, D.~P.~DiVincenzo, J.~A.~Smolin,
  W.~K.~Wootters, \pra {\bf 54}, 3824 (1996).
  
\bibitem{nielsen2000a} Michael~A.~Nielsen and Isaac~L.~Chuang,
  \textit{Quantum Computation and Quantum Information} (Cambridge
  University Press, Cambridge, England, 2000).

\bibitem{toth2004} G.~T\'oth, \pra {\bf 69}, 052327 (2004).

\bibitem{wootters1998} William~K.~Wootters, \prl {\bf 80}, 2245 (1998).

\bibitem{vidal2002} G.~Vidal and R.~F.~Werner, \pra {\bf 65}, 032314 (2002).

\bibitem{ashcroftmermin} Neil~W.~Ashcroft and N.~David~Mermin,
\textit{Solid State Physics},  Philadelphia: Saunders College
(1976).

\bibitem{bose2005} S.~Bose and D.~Home, e-print quant-ph/0505217.
  
\bibitem{cavalcanti2005} D.~Cavalcanti, M.~F.~Santos, M.~O.~Terra
  Cunha, C.~Lunkes, and V.~Vedral, \pra {\bf 72}, 062307 (2005).
  
\bibitem{plenio2005} M.~B.~Plenio, J.~Eisert, J.~Dreissig, and
  M.~Cramer, \prl {\bf 94}, 060503 (2005).

\bibitem{wolf2005} M.~M.~Wolf, e-print quant-ph/0503219.

\end{thebibliography}
\end{document}